\documentclass[runningheads]{llncs}
\usepackage[latin9]{inputenc}
\usepackage{amsmath}
  
\usepackage{graphicx}
\usepackage[noend]{algpseudocode}
\usepackage{algorithmicx,algorithm}
\usepackage{cite}
\usepackage{multirow}
\usepackage{amsmath}  
\usepackage{float}
\usepackage{booktabs} 
\usepackage[noend]{algpseudocode}
\usepackage{algorithmicx,algorithm}
\usepackage{cite}
\usepackage{multirow}
\usepackage{amsmath}  
\usepackage{float}
\usepackage{booktabs} 

\usepackage{graphicx}
\usepackage{subfigure}
\usepackage{bm}
\usepackage{lineno}
\usepackage{url}
\usepackage{tikz}
\usetikzlibrary{positioning}
\usepackage{calc}
\usepackage[colorlinks,linkcolor=black]{hyperref}
\usepackage{subfigure}
\usepackage{bm}
\usepackage{lineno}
\usepackage{url}
\usepackage{tikz}
\usetikzlibrary{positioning}
\usepackage{calc}
\usepackage{diagbox}
\usepackage{url}
\usepackage{verbatim}

\usepackage{graphicx}
\begin{document}
\title{Behavioral Security in Covert Communication Systems}
%
%

\author{Zhongliang Yang \and
Yuting Hu \and
Yongfeng Huang \and
Yujin Zhang
}

\institute{The Department of Electronic Engineering, Tsinghua University, Beijing, 100084, China. \\
\email{\{yangzl15, huyt16\}@mails.tsinghua.edu.cn}\\
\email{\{yfhuang,zhang-yj\}@tsinghua.edu.cn}\\
}
%
%

%
\maketitle              
\begin{abstract}

The purpose of the covert communication system is to implement the communication process without causing third party perception. In order to achieve complete covert communication, two aspects of security issues need to be considered. The first one is to cover up the existence of information, that is, to ensure the content security of information; the second one is to cover up the behavior of transmitting information, that is, to ensure the behavioral security of communication. However, most of the existing information hiding models are based on the ``Prisoners' Model", which only considers the content security of carriers, while ignoring the behavioral security of the sender and receiver. We think that this is incomplete for the security of covert communication. In this paper, we propose a new covert communication framework, which considers both content security and behavioral security in the process of information transmission. In the experimental part, we analyzed a large amount of collected real Twitter data to illustrate the security risks that may be brought to covert communication if we only consider content security and neglect behavioral security. Finally, we designed a toy experiment, pointing out that in addition to most of the existing content steganography, under the proposed new framework of covert communication, we can also use user's behavior to implement behavioral steganography. We hope this new proposed framework will help researchers to design better covert communication systems.

\end{abstract}
\section{Introduction}

Covert communication system, encryption system and privacy system are three basic information security systems which have been summarized by Claude E. Shannon \cite{shannon1949communication}. These three types of information security systems protect people's information security and privacy in cyberspace from different aspects. Among them, the encryption system mainly encrypts important information, makes the unauthorized people cannot decode and read normally, so as to ensure information security\cite{rivest1978method}. The privacy system mainly controls the access to important information and thus ensures the security of information\cite{beller1993privacy}. These two systems can only guarantee the security of information content, but can not cover up the behavior of transmitting secret information. The most important characteristic of a covert communication system is that it can conceal the fact of transmitting secret information, that is, to complete the communication process without causing suspicion from third parties.

In order to achieve truly effective covert communication and avoid being perceived by third parties, there are two aspects of security issues that need to be considered: the concealment of information content and the concealment of information transmission behavior, which correspond to content security and behavioral security, respectively. For a covert communication process, these two issues are indispensable, and they together ensure the concealment and security of the communication process.

Currently, most covert communication systems are under the framework of ``Prisoners' Model" \cite{Simmons1984The}, which is described in detail as follows. Alice and Bob are two prisoners locked in different cells in the prison. They are planning a jailbreak. They are allowed to communicate with each other, but all communicate information must be reviewed by guard Eve. Once Eve finds that they are transmitting secret information, they will be handed over to a cell with the highest security level where they will never be able to escape from. Therefore, they intend to use covert communication methods to embed secret information into common carriers, such as images \cite{fridrich2009steganography}, voices \cite{yang2018aag}, texts \cite{yang2019rnn}, and then communicating with each other. Such information hiding technology is called steganography. Eve's task is to determine as accurately as possible whether the information they are transmitting contains secret information. The technology she uses is called steganalysis. 

In the past few decades, with the development of technology, the steganography and steganalysis methods under the framework of the ``Prisoners' Model" have achieved rapid development and progress on various carriers \cite{fridrich2009steganography,yang2017sudoku,yang2018aag,yang2019rnn,yang2018ts,yang2018automatically,yang2019real,yang2019ts,yang2018rits,yang2019real}. However, with the development of these steganography and steganalysis techniques, we have noticed the limitations of ``Prisoners' Model" which only focus on content security.

A typical communication system consists of three important components: an information sender, a communication channel and an information receiver \cite{shannon1949communication}. The task of the sender is to generate and send information, which can be in the form of images, voices, texts and so on. The communication channel is to pass the signal generated by the sender to the receiver. The receiver is to receive the information carrier transmitted from the channel and obtain the information therein. Simmons' ``Prisoners' Model" only emphasizes content security in the process of information transmission, but ignores the behavioral security of the sender and the receiver. In the scenario it assumes, it even has a very strong assumption that Alice and Bob are allowed to communicate point-to-point. However, in reality, it might be that Alice and Bob's behavior of establishing such point-to-point communication alone is enough to arouse suspicion from others, thus failing to achieve truly covert communication. Therefore, if we want to realize the truly concealment of transmitting secret information, in addition to ensuring the content security in the communication channel, we should also conside the behavioral security of both sender and receiver of the communication system. 

In this paper, we propose a new security framework of covert communication system, which considers both content security and behavioral security in the process of information transmission. We collected and analyzed the behavioral of a large number of active Twitter users, trying to illustrate the security risks that may be brought to covert communication if we only consider content security and neglect behavioral security. Finally, we designed a toy experiment, pointing out that in addition to most of the existing content steganography, under the proposed new framework of covert communication, we can even only rely on user behavior to achieve behavioral steganography, which may bring new ideas and methods to the future covert communication.

In the remainder of this paper, Section II introduces several related works. Section III describes the proposed framework in detail. In section IV, we analyzed the behavioral of a large number of active Twitter users and conducted a series of analytical experiments. Finally, conclusions are drawn in Section V.

\section{Related Works}

\subsection{Steganography and Steganalysis under the ``Prisoners' Model"}

Previous information hiding methods which under the framework of the ``Prisoners' Model"\cite{Simmons1984The} mainly focuse on content security, they try to find the best way to hide secret information into common carriers. These steganographic methods can be divided into different types according to different kinds of carrier, like image steganography \cite{fridrich2009steganography}, text steganography \cite{yang2019rnn}, audio steganography \cite{yang2018aag} and so on. In addition, according to different steganography means, they can also be classified to modification-based steganography \cite{fridrich2009steganography} and generation-based steganography \cite{yang2019rnn,yang2018aag}.

Figure 1 shows the overall framework of the ``Prisoners' Model" and we can model it in the following mathematical form. Assuming that there are three spaces: carrier space $\mathcal{C}$, key space $\mathcal{K}$, and secret message space $\mathcal{M}$. The process of information hiding can be represented by the function $f()$. If Alice adopts a steganographic method based on the modification mode, she selects a common carrier $c$ from the carrier space $\mathcal{C}$, which can be an image, voice or text. Then, under the control of the secret key $k_A$ from the key space $\mathcal{K}$, Alice embeds the secret information $m$ into the carrier $c$ by modifying it, that is:

\begin{equation}
Emb: \mathcal{C} \times \mathcal{K} \times \mathcal{M} \to \mathcal{S}, f_{mod}(c,k_A,m) = s.
\end{equation}

\noindent Correspondingly, if Alice adopts the steganographic method based on the carrier automatic generation, she does not need to be given a carrier in advance. She can automatically generate a steganographic carrier $s$ according to the secret information $m$ that needs to be transmitted, that is:

\begin{figure}[!tp]
\centering
\includegraphics[width=\linewidth]{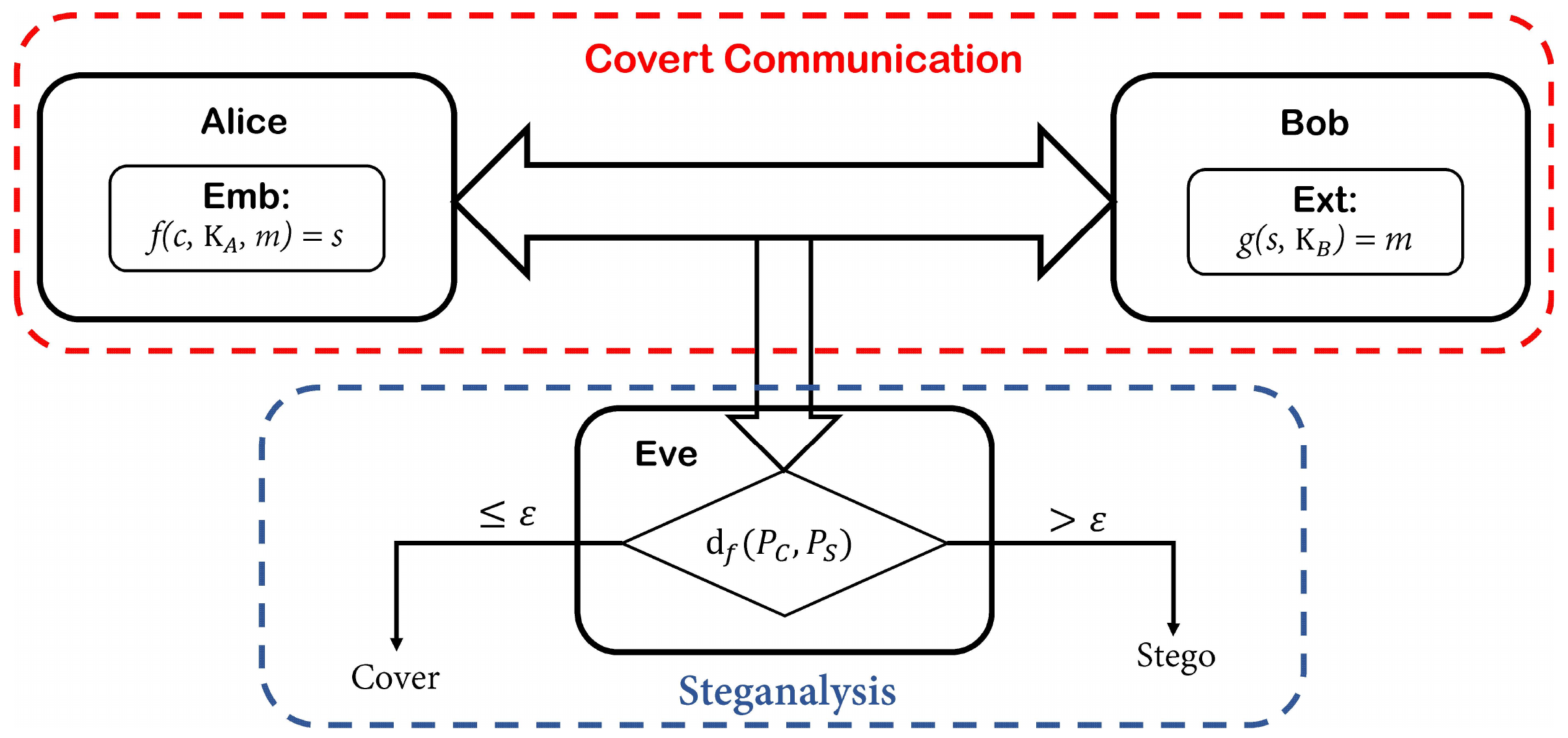}
\caption{Steganography and steganalysis within the framework of ``Prisoners' Model".}
\label{fig:2}
\end{figure}

\begin{equation}
Emb: \mathcal{K} \times \mathcal{M} \to \mathcal{S}, f_{gen}(k_A,m) = s.
\end{equation}

When Bob receives the carrier $s$ containing secret information from Alice, under the control of the decoding key $k_B$, he can extract the embedded secret message $m$ from the steganographic carrier $s$ by using the corresponding extraction function $g()$, that is:

\begin{equation}
Ext: \mathcal{S} \times \mathcal{K} \to \mathcal{M}, g(s,k_B) = m.
\end{equation}

In order not to arouse suspicion of Eve and ensure the security of secret information, according to Cachin's analysis of steganographic security \cite{cachin2004information}, Alice and Bob should try their best to reduce the statistical distribution differences between normal carriers and steganographic carriers, that is:

\begin{equation}
d_f(P_{\mathcal{C}},P_{\mathcal{S}}) \leq \varepsilon_f .
\end{equation}

\noindent Among them, $P_{\mathcal{C}}$ and $P_{\mathcal{S}}$ represent the statistical distribution of normal carriers and steganographic carriers respectively. $\varepsilon_f$ is a number greater than 0, which can be used to measure the concealment of steganographic algorithm $f()$. For Eve, her job is to analyze and judge whether the information carrier being transmitted has deviated from the statistical distribution of the normal carriers, and then determine whether the carrier contains secret information.

The limitation of ``Prisoners' Model" is that it only considers the content security in the communication process and ignores the behavioral security of the communication ends, which we think is incomplete for achieving truly effective covert communication.

\subsection{Behavioral Concealment and Behavioral Analysis}

At present, most covert communication models mainly focus on the content security of the information transmitted in the intermediate channel, but in fact, both ends of the communication system can also participate in the entire covert communication process. For example, during World War II, some people used several specific behaviors agreed in advance to convey covert information \cite{li2018lost}. In recent years, some researchers have begun to study how to use specific online behaviors to convey covert information. For example, N. Pantic \emph{et al.} \cite{pantic2015covert} proposed a steganographic method that represents different secret information by controlling the length of Twitter published. X. Zhang \cite{zhang2017behavior} suggested that different secret messages could be conveyed by giving ``love" marks to the information on social media. S. Li \emph{et al.} \cite{li2018lost} defined a ``bits to activities" mapping algorithm and tried to use different online behaviors to convey secret message.

These covert communication methods can be implemented smoothly because they utilize the blind area of current steganalysis methods under the ``Prisoners' Model" framework, that is, the lack of modeling and analysis of user's behavior. However, it is noteworthy that in recent years, with the development of technology, more and more methods for detecting abnormal behaviors on the Internet have emerged\cite{costa2017modeling,chino2017voltime,ranshous2015anomaly,anand2017anomaly}. In this case, if we still only consider the content security in the covert communication process, without considering the behavior security, it will bring a great security risk to the entire covert communication system no matter how concealment the secret information is embedded.

\section{The Proposed Framework}

In order to describe the proposed framework more conveniently, we first describe a virtual scene. We assume that Alice and Bob are two intelligence personnel disguised as ordinary people. Their job is to gather intelligence. They have sneaked into two different target hostile areas and each of them have collected some intelligence information. Now they need to communicate with each other, exchange the information they have acquired and verify them with each other in order to make the best decisions. However, due to previous actions, these two areas have been suspected by the enemy, which can be collectively refered as Eve. Eve suspects that Alice and Bob have infiltrated these two areas, but she is not sure who they are. She was authorized to review all the communications between these two regions in the hope of finding Alice and Bob. According to Kerckhoffs's principle\cite{shannon1949communication}, we can expand Eve's capabilities as much as possible. We assume that Eve can get the content of each communication between any two people in these two regions, so she can perform steganalysis by analyzing the statistical distribution characteristics of the communication content. In addition, she can also know the communication behavior between any two people in these two regions, so she can also perform steganalysis by analyzing the communication behavior. We should also assume that Eve is familiar with all kinds of steganographic algorithms that Alice and Bob may adopt, and only does not know the specific steganography parameters (i.e. the secret key) that Alice and Bob adopt.

For Alice and Bob, in order to successfully accomplish the task and ensure their own safety, they need to consider two aspects of security. Firstly, they need to ensure that the information they transmit is well concealed and not easily to be detected; secondly, they should behave as normal as possible and thus not expose themselves. They first ruled out the use of encryption methods to transmit information, because the transmission of encrypted information in the public channel will likely arouse suspicions. They eventually choose to use steganographic methods for covert communication. But only assurance the content concealment of each point-to-point communication as required by the ``Prisoners' Model" is obviously not enough. They need to disguise themselves as normal people in a public communication network, cover up their behavior of exchanging secret information, and ensure the content security of each communication at the same time, thus they can achieve real covert communication.

The framework of the whole scenario is shown in Figure 2 and the mathematical descriptions for this scenario and task are as follows. We define the entire public communication network as a graph series $\textbf{G} = \{G_t\}^T_{t=1}$, where $G_t$ denotes the graph at the $t$-th moment. Graph $G_t$ of time step $t$ is composed of vertices set $V_t$ and edges set $E_t$, that is:

\begin{equation}
\begin{aligned}
&\textbf{G} = \{G_t\}^T_{t=1}. &s.t. \quad \forall t\in [1,T], G_t = \{V_t,E_t \subseteq (V_t \times V_t) \}.
\end{aligned}
\end{equation}

\noindent The vertices in $G_t$ represent the users in the social network and edges represent the connections between users:

\begin{equation}
\begin{aligned}
&\forall i,j\in [1,N], \forall t\in [1,T],\\
&V_t = \cup^N_{n=1}\{v^i_t\}, \quad E_t = \cup\{e^{i,j}_t\}|_{i,j \in [1,N]}, \quad e^{i,j}_t: v^i_t \to v^j_t.
\end{aligned}
\end{equation}

\begin{figure*}[!tp]
\centering
\includegraphics[width=\linewidth]{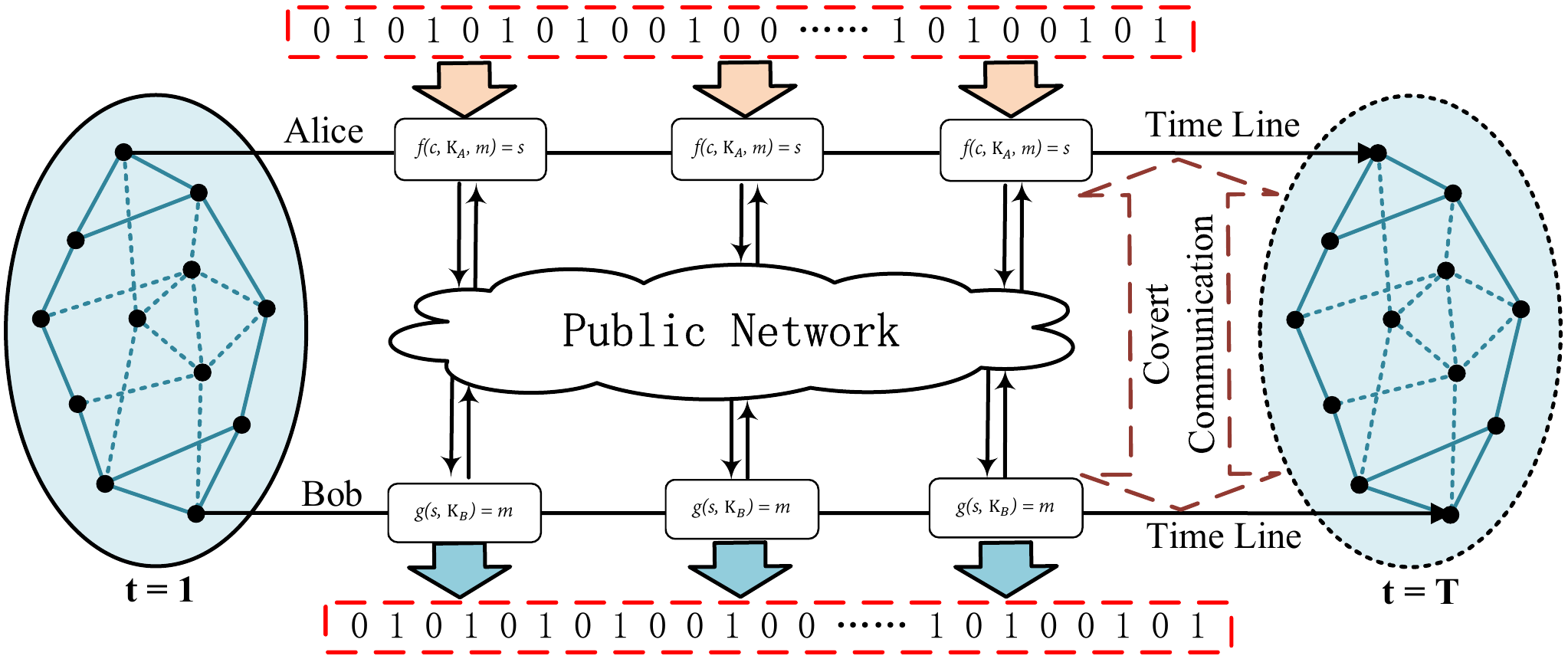}
\caption{Covert communication in public cyberspace.}
\label{fig:2}
\end{figure*}

We can represent Alice and Bob in this public network as $v^A$ and $v^B$. Without affecting the nature of this problem, for the convenience of the following discussion, we assume that all the secret information is sent by Alice and extracted by Bob. According to the previous description, in order to achieve real covert communication, we need to consider both content security and behavioral security in the communication process. At the same time, we notice from Figure 2 that the content and behavior of each user in the network have instantaneous and temporal characteristics. Next we will discuss and model them separately. 

\begin{itemize}

\item \textbf{Instantaneous Content Security}

In order to ensure instantaneous content security, Alice and Bob only need to ensure that at each time, the difference in the statistical distribution characteristics of the carrier they are transmitting and the normal carriers is small enough. This is very similar to formula (1) - (4). We assume that at time step $t$, Alice needs to pass a covert message $m_t$ to Bob. Alice selects a carrier $c_t$ from the carrier space $\mathcal{C}$, uses a steganographic function $f_t()$ to embed the secret information $m_t$ into common carrier $c_t$ under the guidance of the secret key $k^A_t \in \mathcal{K}$, and gets the steganographic carrier $s_t$. To minimize the impact on the statistical features distributions of carriers, so as to avoid arousing suspicion of other people about the transmission content, the steganographic function $f_t()$ needs to satisfy the following constraints:

\begin{equation}
\begin{aligned}
&\forall t\in [1,T],
& \left\{\begin{array}{l}
f_t(c_t,k^A_t,m_t) = s_t, \\
g_t(s_t,k^B_t) = m_t,\\
d_1(P_\mathcal{C},P_\mathcal{S}) \leq \varepsilon_1.\\
\end{array}  
        \right.
\end{aligned}
\end{equation}

\noindent Where $g_t()$ and $k^B_t$ are the extraction function and the extraction key corresponding to $f_t()$ and $k^A_t$, respectively. $P_\mathcal{C}$ and $P_\mathcal{S}$ represent the statistical distribution characteristics of the normal carriers and the steganographic carriers, respectively, and $d_1()$ is a measurement function for measuring the statistical distribution difference between the normal carriers and the steganographic carriers. $\varepsilon_1$ is a value greater than 0 and it measures the security of the steganographic function $f_t()$. The smaller the value of $\varepsilon_1$, the stronger the concealment of $f_t()$.

\item \textbf{Temporal Content Security}

However, it is not enough to only guarantee the content security of a single communication. For users in social networks, according to their own interests and characteristics, usually the contents they publish or pay attention to have a certain relevance. Therefore, for Alice and Bob, the content they upload and download should also have a certain temporal correlation. For example, if a person is a big fan of basketball and his social media content in the past short term is sports-related, then we have reason to guess that the next content may also be sports-related. Or if a person always publishes negative content on social media for a long time, we can also estimate that the emotions contained in the following content may be negative. Conversely, if a social account publishes multiple messages in a short period of time (e.g. within an hour) and the topic changes frequently, or emotional switching frequently, we can at least suspect that the account is unusually. We use $c^i_{t_1:t_2}$ to represent the sequence of information published or downloaded by user $v^i$ during the period from $t_1$ to $t_2$. Suppose $H()$ is a function which can measure the temporal relevance of sequence information. Therefore, in order to achieve temporal content security, Alice and Bob should ensure:

\begin{equation}
\begin{aligned}
&\forall t_1, t_2 \in [1,T], t_1<t_2,
& \left\{\begin{array}{l}
d_2(H(s^A_{t_1:t_2}),\overline{H(c_{t_1:t_2})}) \leq \varepsilon_2.\\
d_2(H(s^B_{t_1:t_2}),\overline{H(c_{t_1:t_2})}) \leq \varepsilon_2.\\
\end{array}  
        \right.
\end{aligned}
\end{equation}

\noindent Where $\overline{H(c_{t_1:t_2})}$ represents the average score of a large number of sequence social information published by normal users. $d_2()$ reflects the difference between temporal relevance score of sequence social information published by Alice and normal users. $\varepsilon_2$ is a value greater than 0, the smaller it is, the stronger the concealment.

\item \textbf{Instantaneous Behavioral Security}

For each moment $t \in [1,T]$, the behavioral security of each user $v^i_t$ in cyberspace can be divided into two aspects: one is the behavioral security of himself, the other is the contacts with the people around him, that is, the behavioral security of a single vertex and the edges around it. Extreme examples such like Bob only downloads information published by Alice, or Alice publishs an unusually large amount of content in a very short time (for example, 100 images in one minute), may bring risks of suspicion. Here, we define two behavior scoring functions $I()$ and $J()$ to analyze the behavior of vertexes and edges in the public network, respectively. Then to achieve instantaneous behavioral security, Alice and Bob's behavior at each moment needs to meet the following constraints:

\begin{equation}
\begin{aligned}
& \left\{\begin{array}{l}
\forall t\in [1,T],\\
d_3(I(v^A_t), \overline{I(v_t)}) \leq \varepsilon_3, \quad d_3(I(v^B_t), \overline{I(v_t)}) \leq \varepsilon_3,\\
d_4(J(E^A_t), \overline{J(E_t)}) \leq \varepsilon_4, \quad d_4(J(E^B_t), \overline{J(E_t)}) \leq \varepsilon_4.\\
\end{array}  
        \right.
\end{aligned}
\end{equation}

\noindent Where $E^A_t$ and $E^B_t$ represents all the edge sets associated with Alice and Bob at $t$-th time step, and $\overline{I(v_t)}$ and $\overline{J(E_t)}$ represent the average behavior score of the normal users and their connections, respectively. For instantaneous behavioral security, the statistical distribution difference between Alice's and Bob's online behavior and that of normal users' behavior should be less than a threshold $\varepsilon_3$ and $\varepsilon_4$.

\item \textbf{Temporal Behavioral Security}

A.F.Costa \emph{et al.} \cite{costa2017modeling} and Y.T.Daniel \emph{et al.} \cite{chino2017voltime} have found that for many ordinary users, when logging in and using these social media, their behavioral records have a significant time distribution. This shows that users may have their own habits of using these social media and these behavior on social media is likely to have a temporal correlation. For example, some people are accustomed to watching a video on social media before going to bed, or browsing their friends' information and commenting on social media after getting up every day. From the perspective of behavioral security, it would be much less likely for Alice and Bob to expose themselves if they found a large number of normal users' usage habits and statistical characteristics and then imitated their usage habits. The temporal characteristics of users' online behavior can also be divided into two aspects: vertex behavior and edge behavior, then the constraints should be:

\begin{equation}
\begin{aligned}
&\left\{\begin{array}{l}
\forall t_1, t_2 \in [1,T], t_1<t_2,\\
d_5(H(v^A_{t_1:t_2}),\overline{H(v_{t_1:t_2})}) \leq \varepsilon_5,\quad d_5(H(v^B_{t_1:t_2}),\overline{H(v_{t_1:t_2})}) \leq \varepsilon_5.\\
d_6(H(E^A_{t_1:t_2}),\overline{H(E_{t_1:t_2})}) \leq \varepsilon_6, \quad d_6(H(E^B_{t_1:t_2}),\overline{H(E_{t_1:t_2})}) \leq \varepsilon_6.\\
\end{array}  
        \right.
\end{aligned}
\end{equation}

\end{itemize}

\noindent Where $H(v^A_{t_1:t_2})$, $H(v^B_{t_1:t_2})$ and $H(E^A_{t_1:t_2})$, $H(E^B_{t_1:t_2})$ represent the temporal relevance of their sequential behavior of Alice and Bob themselves and their interaction with the people around them, respectively. $\overline{H(v_{t_1:t_2})}$ and $\overline{H(E_{t_1:t_2})}$ represent the corresponding distribution characteristics of ordinary users. To ensure temporal behavioral security, the difference of their distribution should be less than the threshold $\varepsilon_5$ and $\varepsilon_6$.

Based on the above analysis, the overall framework for covert communication in public networks has been summarised in Figure 3, and in order to achieve real covert communication, the security constraints that Alice and Bob need to meet are shown in Table 1.

\begin{figure*}[!tp]
\centering
\includegraphics[width=\linewidth]{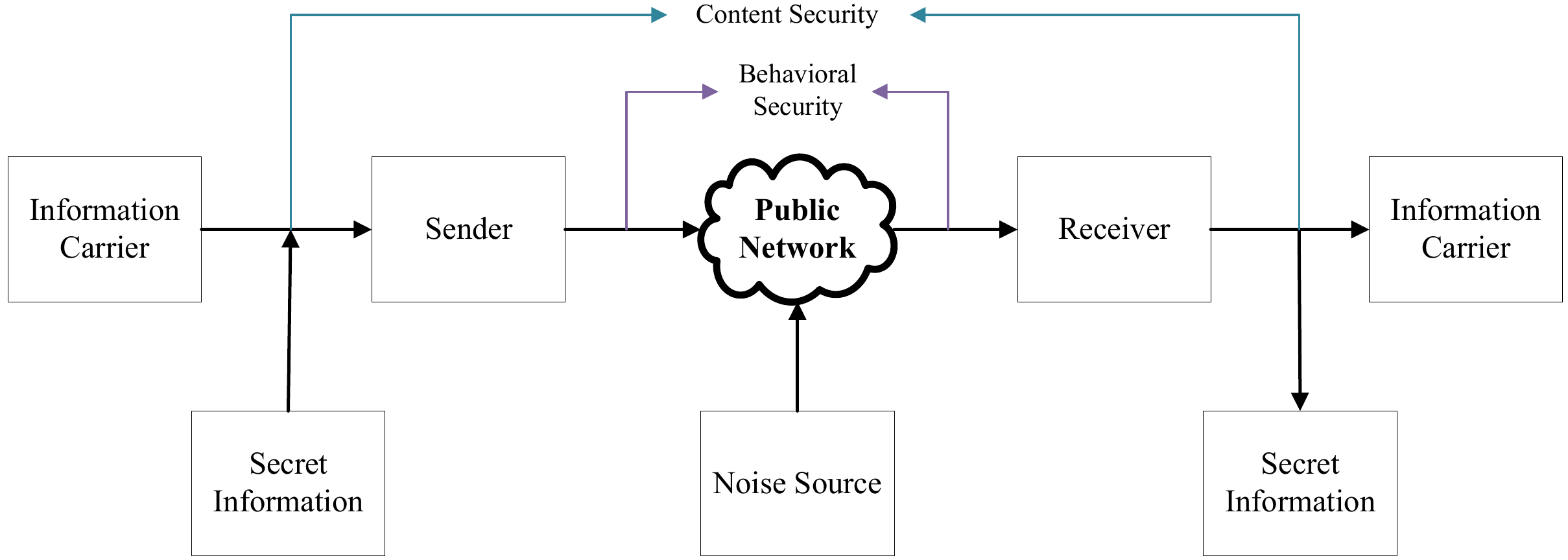}
\caption{The overall framework for covert communication in public networks.}
\label{fig:2}
\end{figure*}

\begin{figure*}[ht]
\centering
\includegraphics[width=\linewidth]{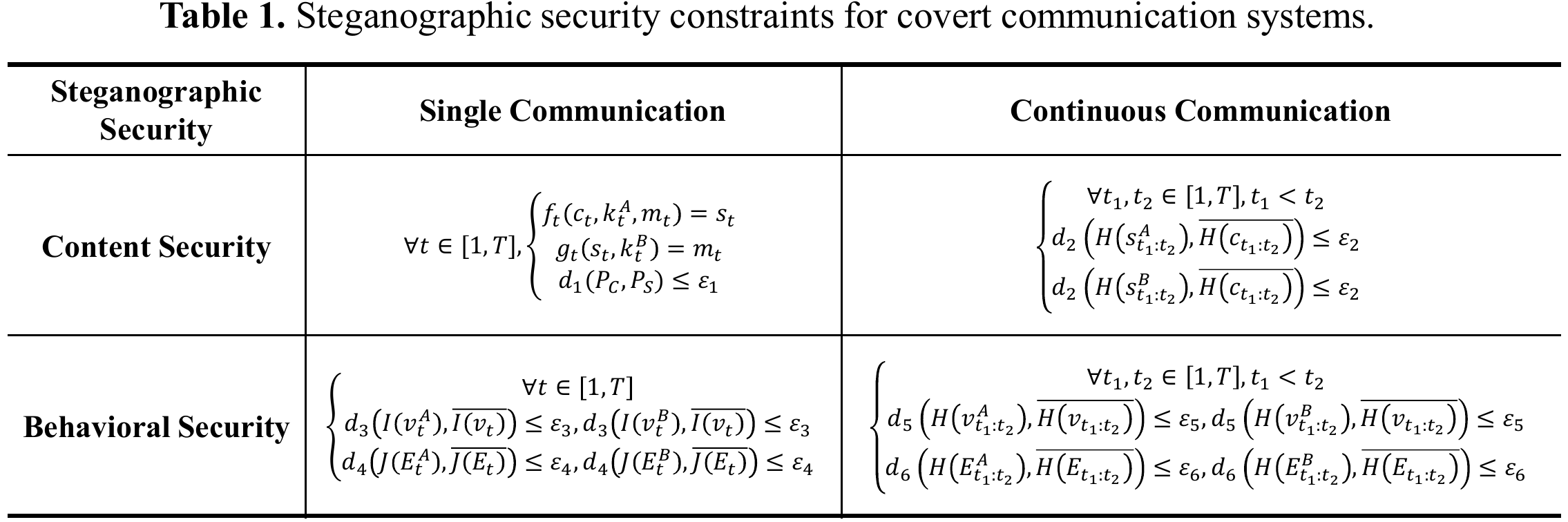}
\label{fig:2}
\end{figure*}

\section{Experiments and Analysis}

In this section, we first analyzed the crawled Twitter user behavior and content characteristics from some aspects. And then we analyze the possible risks if we only consider content security without considering behavior security. Finally, we designed a toy experiment, pointing out that in addition to most of the existing content steganography, in the proposed new framework of covert communication, we can even only rely on user behavior to achieve behavioral steganography.

\subsection{Data Collection and Analysis}

According to the previous section, in order to achieve truly effective covert communication, Alice and Bob need to cover up themselves from both content and behavior aspects. To achieve this goal, we first need to analyze the content and behavioral characteristics of a large number of normal users on public networks. In this work, we crawled 317,375 tweets from 1,147 active Twitter users (defined as publish Twitter messages more than 90 in a month) for the period from 2019-06-15 00:00:00 to 2019-07-15 23:59:59. Some detail information about these users and tweets can be found in Table 2.

\setcounter{table}{1}

\begin{table}[ht]
\renewcommand\arraystretch{1.4}
  \centering
  \caption{Some detail information about the crawled twitter users and tweets}
  \setlength{\tabcolsep}{2mm}{
  \begin{tabular}{c|c|c|c}
    \toprule[1.5pt]
    User Number &Tweet Number &Rate of retweet &Average forwarding\\
    \hline
    1,147 &317,375 &34.55\%  &151.0\\
    \hline
    \hline
    Average comments &Average follower &Average following\\
    \hline
    1871.7 &23,036 &8761.2\\
    \bottomrule[1.5pt] 
  \end{tabular}}
  
  \label{tab 1}
\end{table}

\begin{figure}[ht]
\centering

\subfigure[]{
\begin{minipage}[t]{0.43\linewidth}
\centering
\includegraphics[width=\linewidth]{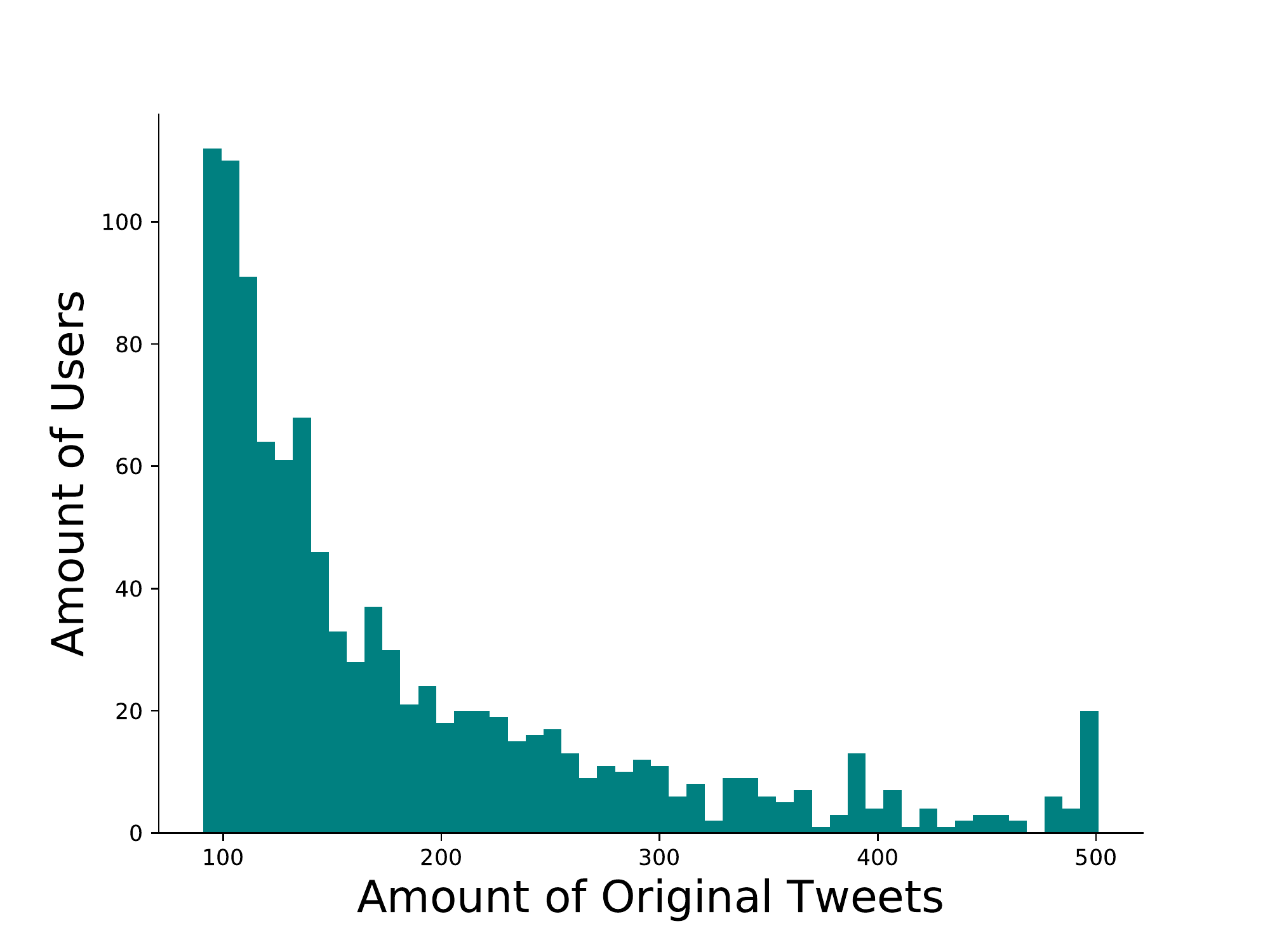}
\end{minipage}%
}%
\subfigure[]{
\begin{minipage}[t]{0.43\linewidth}
\centering
\includegraphics[width=\linewidth]{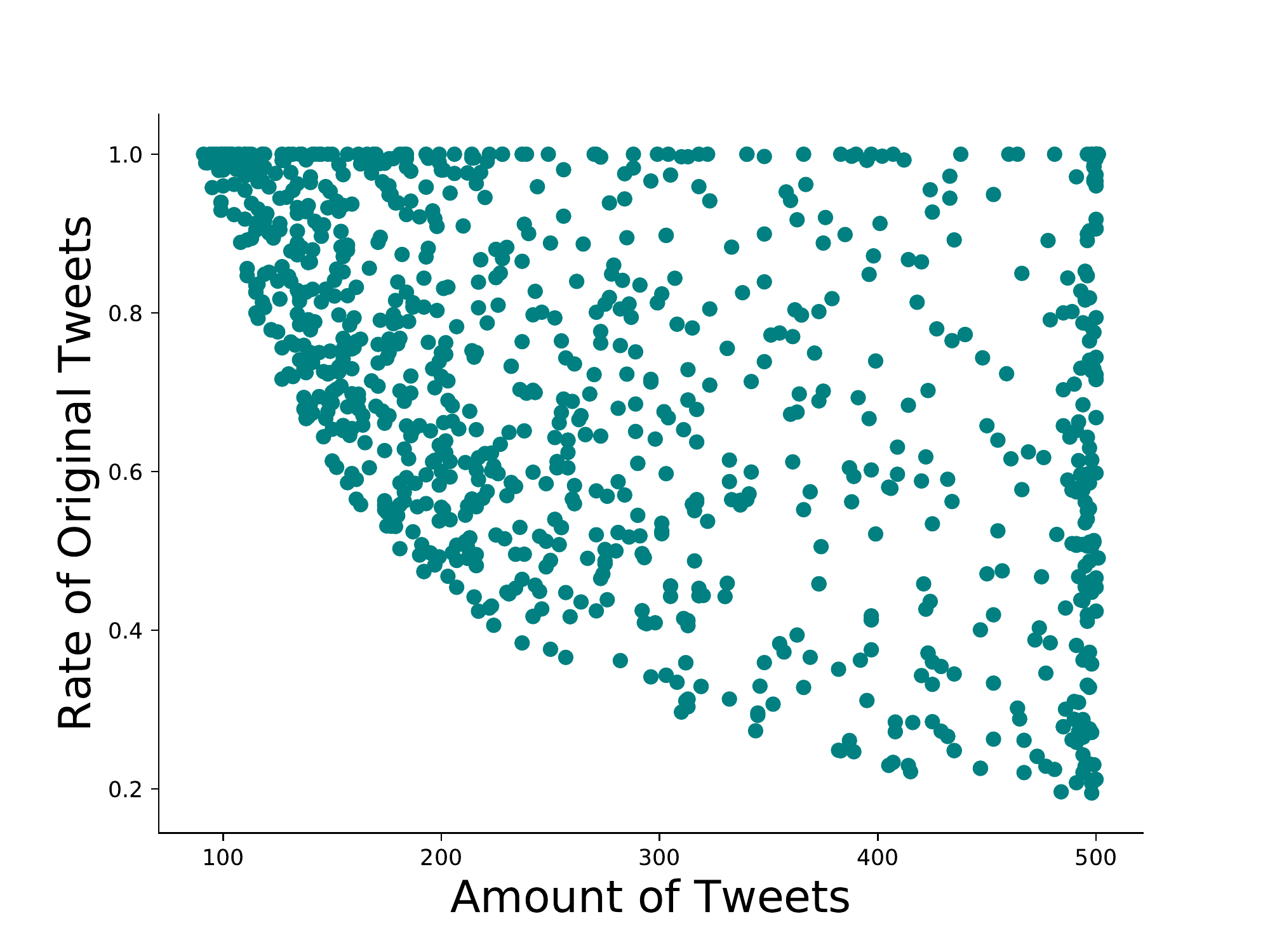}
\end{minipage}%
}%
\quad
\subfigure[]{
\begin{minipage}[t]{0.43\linewidth}
\centering
\includegraphics[width=\linewidth]{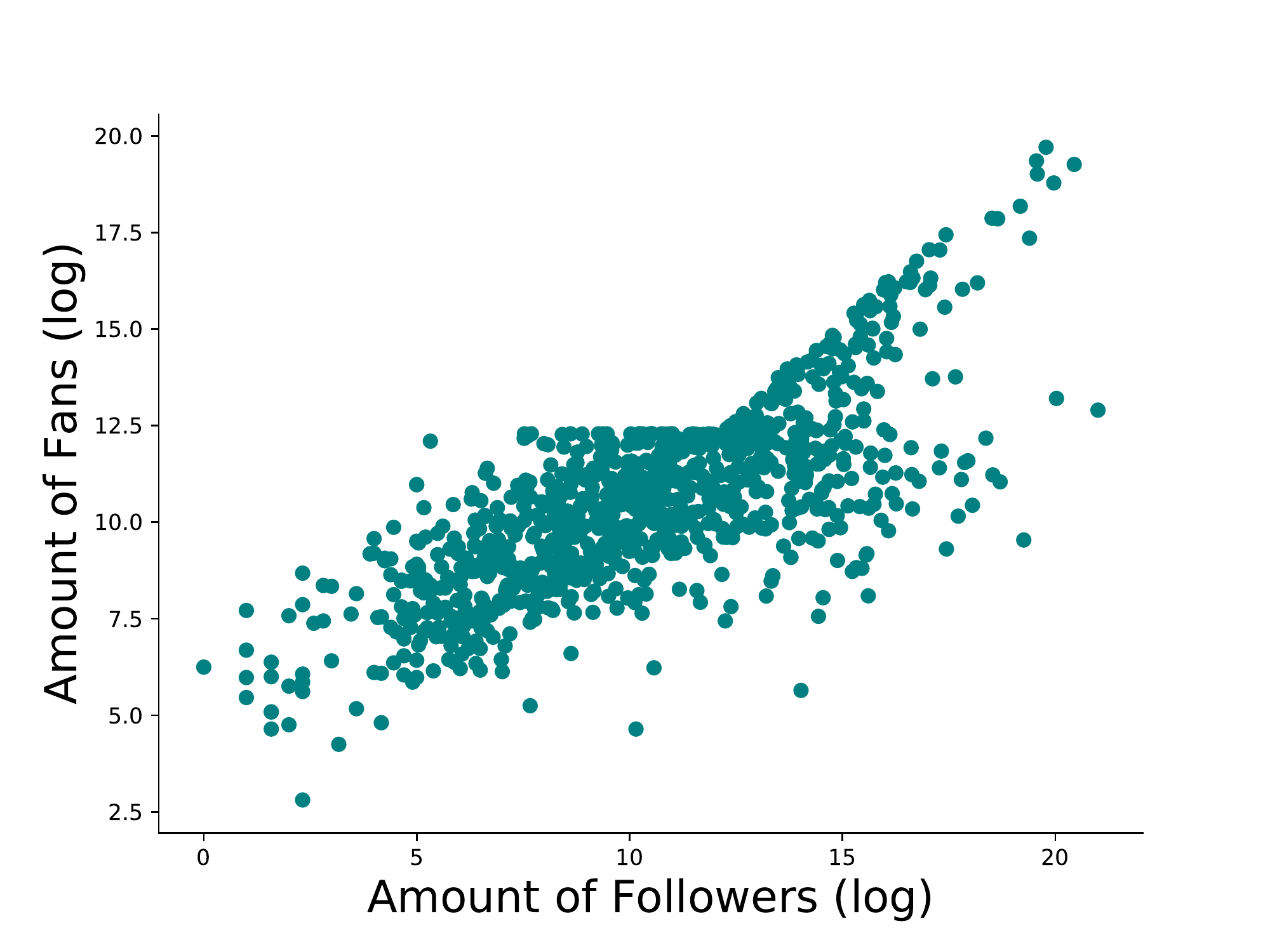}
\end{minipage}%
}%
\subfigure[]{
\begin{minipage}[t]{0.43\linewidth}
\centering
\includegraphics[width=\linewidth]{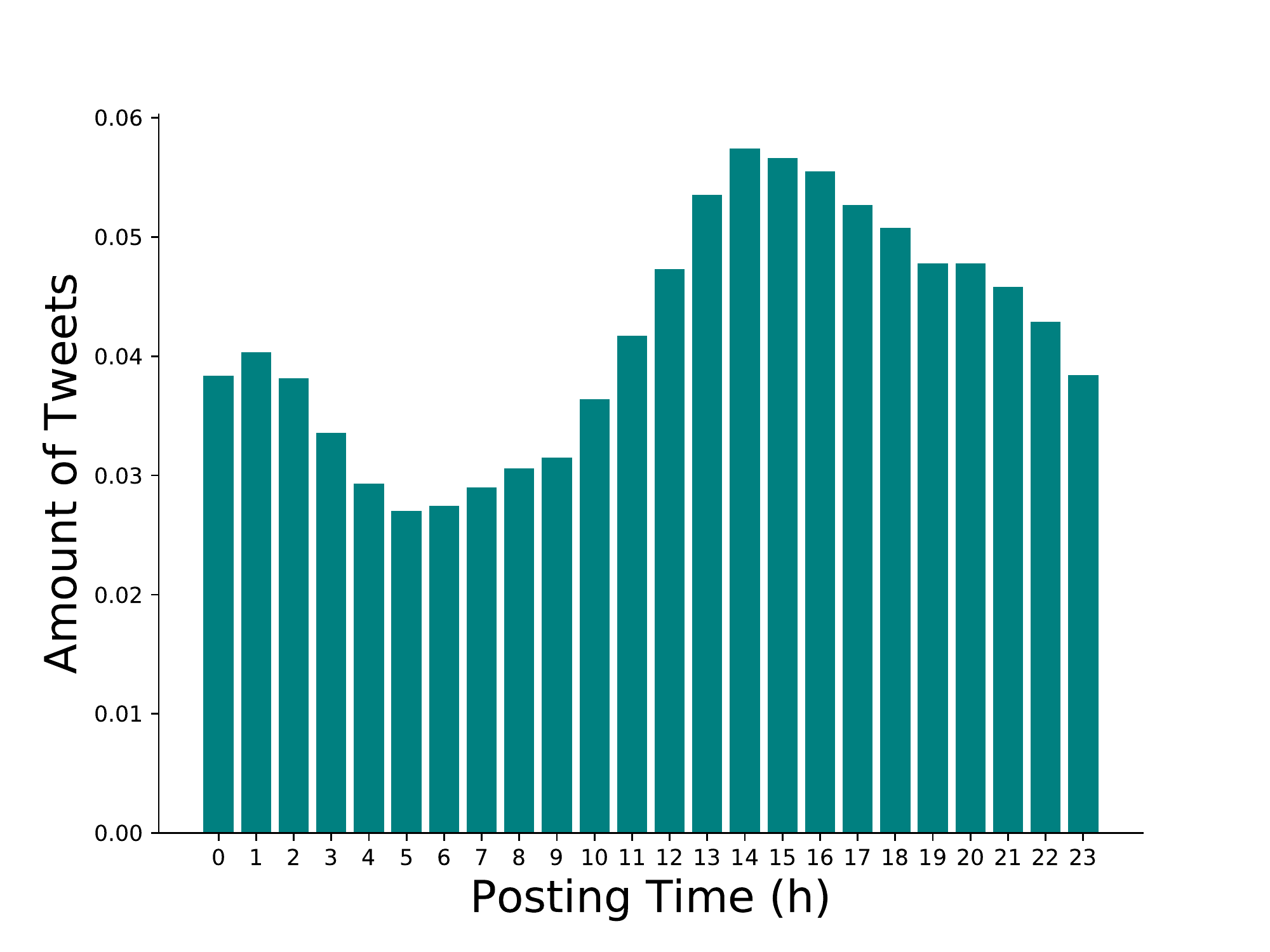}
\end{minipage}%
}%
\centering
\caption{Some statistical distribution characteristics of online behaviors of active twitter users.}
\end{figure}

We made some statistical analyses of these crawled Tweets, and the results are shown in Figure 4. From these results, we can at least see that when these unrelated users publish content on social networks, their behavior will show a regular distribution on the whole, rather than disorder. For example, Figure 4 (b) reflects the relationship between the total number of Twitters and the rate of original Twitters published by these active users in a month. It forms a very regular fan-shaped distribution. If Alice ignores these statistical rules, Twitter publishing falls outside the sector. Even if the content she publishes is normal, it may be recognized as abnormal behavior and expose herself. Figure 4(d) shows the statistical distribution of the average number of Twitters posted by users within 24 hours a day. They also form a very regular distribution. If the time when Alice publishes information does not coincide with that of most people, such as always publishing information at the lowest point of normal people's probability of republishingleasing information, it may also be considered abnormal.

In addition, from the perspective of content security, we analyzed in Section III that social information published by users (especially active users) is likely to have temporal content relevance. We use the VADER model proposed in \cite{hutto2014vader} to analyze the sentiment of crawling Twitters. The sentiment score of each Twitter is distributed between [-1,1]. The closer the value is to 1, the more positive it is, and the closer it is to -1, the more negative it is. Then, we use permutation entropy\cite{bandt2002permutation} to calculate the random degree of Twitter sentiment values of these active users within a month. As a measure, permutation entropy (PE) is widely used in the analysis, prediction and detection tasks of time series. The value of PE (in a range of [0,1]) represents the degree of randomness or predictability of the time series, the closer to 0, the more regular the sequence is.

In contrast, we have also constructed 100 virtual users and used recently appeared steganographic text generation algorithm \cite{yang2019rnn} to generate 100 steganographic twitters for each virtual user. Then we use the same method to calculate the sentiment value of each steganographic Twitter and then use permutation entropy to calculate its random degree. We use different embedding rates (bits per word) to generate multiple sets of Twitters, and the final test results are shown in Figure 5.

\begin{figure}[ht]
\centering
\includegraphics[width=\linewidth]{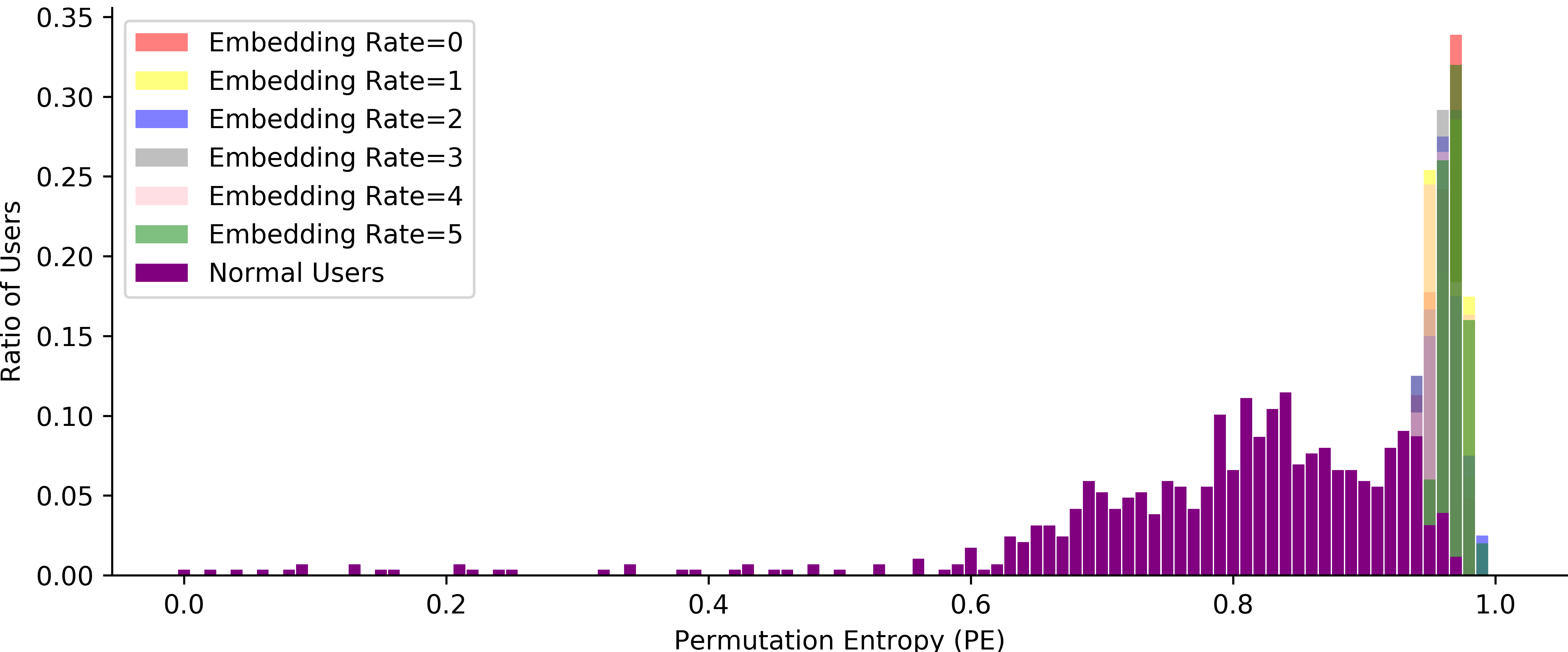}
\caption{The permutation entropy distribution of sentiment values of Twitter published by normal users within one month and steganographic Twitter generated by model proposed in \cite{yang2019rnn}.}
\label{fig:2}
\end{figure}

As can be seen from Figure 5, the PE distribution of steganographic Twitter (regardless of embedding rate) generated is very close to 1, which means that the sentiment values of these Twitters are almost random sequences. In contrast, the PE distribution of real Twitter published by normal active users is closer to 0, indicating that the real user generated Twitter sequence has a certain predictability of sentiment state. This further validates our previous analysis that no matter how concealment the steganographic samples generated by Alice each time, once the temporal correlation is ignored, it is very likely to expose itself.

These analyses show that a large number of normal users' online content and behavior have certain statistical characteristics, and neglecting any aspect of them will bring security risks, so that effective covert communication cannot be realized. Therefore, in order to achieve real covert communication, whether single communication or continuous communication, we need to consider both content security and behavioral security to ensure the concealment of communication.

\subsection{Behavioral Steganography Based on Posting Time}

In fact, by effectively expanding the ``Prisoners' Model", on the one hand, it brings greater challenges to Alice and Bob, but on the other hand, it also brings them new ideas of steganography. They may even not need to modify the content of the original carrier, only use the online behavior to transmit secret information, which can be called behavioral steganography. In this section, we will present a very simple but effective method, to illustrate how can we conduct behavioral steganography.

The steganography models based on the ``Prisoners' Model" usually need to analyze the statistical distribution characteristics of the normal carrier, and then modify the insensitive features of it and thus to embed secret information. This kind of methods can be called content steganography. Similarly, in order to implement behavioral steganography, we first need to analyze the statistical distribution characteristics of a large number of normal users' behaviors.

Figure 4(d) shows the statistical distribution of the posting time within 24 hours a day. Therefore, we can consider using the behavior of posting time to transmit secret information. For example, if Alice divides each day into hours, she can get 24 time periods for sending information and also the probability of sending information in each time hour according to Figure 4(d), that is :

\begin{equation}
\begin{aligned}
& \left\{\begin{array}{l}
\{h_1,h_2,...,h_{24}\},\\
\{P(h_1),P(h_2),...,P(h_{24})\}.\\
\end{array}  
        \right.
\end{aligned}
\end{equation}

Then Alice can construct a Huffman tree and encode each time period according to the corresponding probability. Each leaf node represents a period of time, and the encoding rule can be 0 on the left and 1 on the right. The Huffman coding for each time period can be find in Table 3.

\begin{table}[ht]
\centering
\renewcommand\arraystretch{1.4}
\caption{\label{tab:1}Huffman code table for each posting time during one day.}
\begin{tabular}{c|c|c|c|c|c|c|c|c|c|c|c}
\toprule[1.5pt]
O'clock &Code &O'clock &Code &O'clock &Code &O'clock &Code &O'clock &Code &O'clock &Code\\
\hline
0-1 &11001 &1-2 &11011 &2-3 &11000 &3-4 &10110 &4-5 &10011 &5-6 &01010\\
\hline
6-7 &01011 &7-8 &10010 &8-9 &10100 &9-10 &10101 &10-11 &10111 &11-12 &11100\\
\hline
12-13 &11111 &13-14 &0100 &14-15 &1000 &15-16 &0111 &16-17 &0110 &17-18 &0011\\
\hline
18-19 &0010 &19-20 &0000 &20-21 &0001 &21-22 &11110 &22-23 &11101 &23-24 &11010\\
\bottomrule[1.5pt] 
\end{tabular}
\end{table}

Then each time according to the secret bit stream, Alice searches from the root node of Huffman tree until the corresponding leaf node is found, indicating that the steganographic information should be sent out in the corresponding time. We simulated Alice to transmit secret information at a specified time using the above method and embed random bit streams into the sending time. We simulated Alice sending 5,000 messages, and then we counted the distribution of these sending times, results have been shown in Figures 6. From figure 6, we can see that for this steganographic method, while transmitting hidden information, Alice can still obey the statistical characteristics of normal users' behavior, so it can achieve a certain degree of behavioral security. At the same time, it is worth noting that Alice has not made any changes to the carrier content, so the content security can also be guaranteed.

\begin{figure}[ht]
\centering
\includegraphics[width=\linewidth]{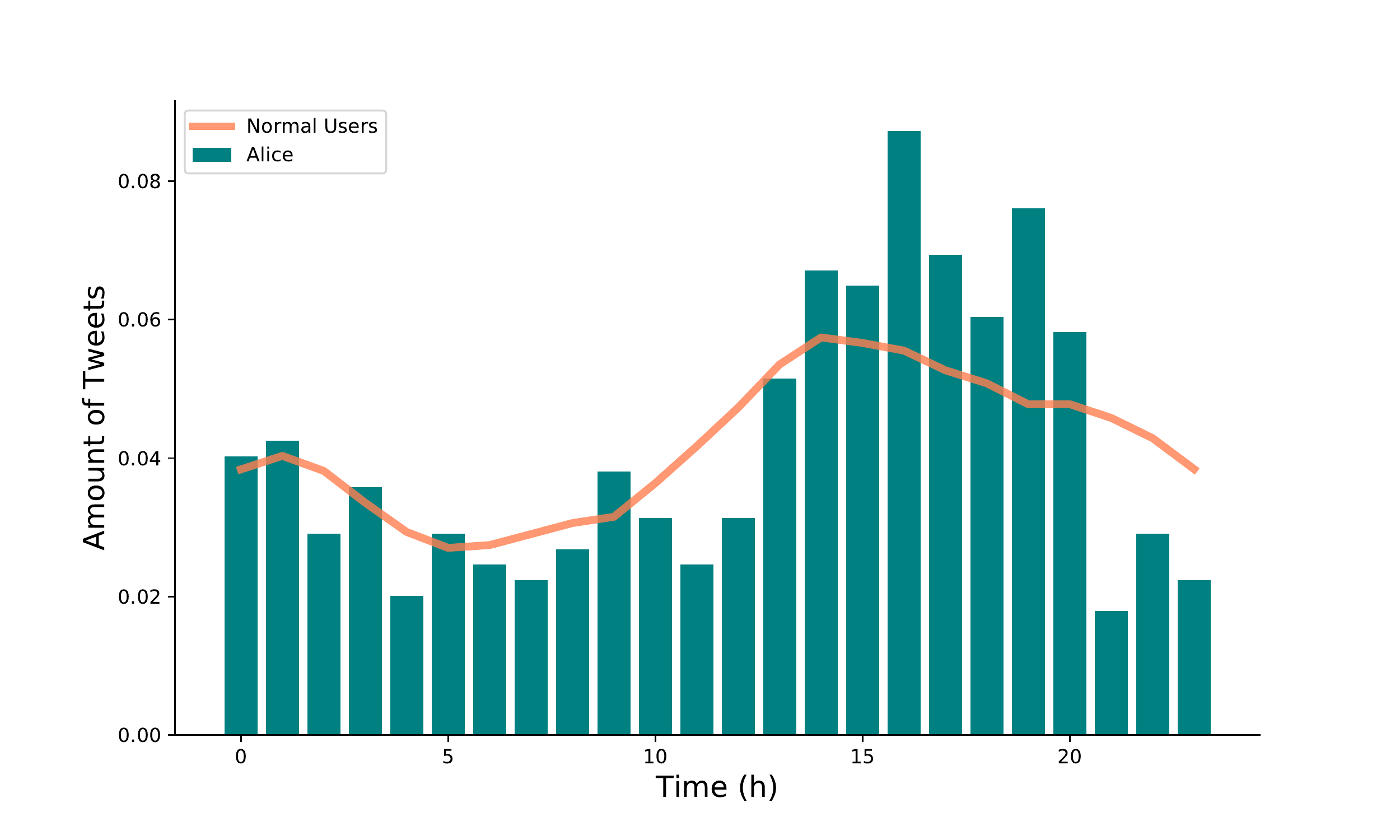}
\caption{The statistical distribution difference between the steganographic behavior and the real behavior after carrying the covert information with the transmission time.}
\label{fig:2}
\end{figure}

Although this is just a simple example of using network behavior to carry covert information. But it proves that we can jump out of the scope of content security, and only use the behavior to achieve covert communication. We think that this may inspire more new types of covert communication technologies and methods in the future.

\section{Conclusion}

In this paper, we propose a new covert communication framework, which considers both content security and behavioral security in the process of information transmission. We give a complete security constraint under the proposed new framework of covert communication. In the experimental part, we use a large amount of collected real Twitter data to illustrate the security risks that may be brought to covert communication if we only consider content security and neglect behavioral security. Finally, we designed a toy experiment, pointing out that in addition to most of the existing content steganography, in the proposed new framework of covert communication, we can even only rely on user behavior to achieve behavioral steganography. We think this new proposed framework is an important development to the current covert communication system. We hope that this paper will serve as a reference guide for researchers to facilitate the design and implementation of better covert communication systems.

\section*{Acknowledgment}

The authors thank Dr. Shujun Li for constructive communications. This work was supported in part by the National Key Research and Development Program of China under Grant SQ2018YGX210002 and the National Natural Science Foundation of China (No.U1536207, No.U1705261 and No.U1636113).


%
%
%
%
\bibliographystyle{IEEEtran}
\bibliography{IEEEexample}
\end{document}